\documentclass[cits]{PoS}

\newcommand{\be}{\begin{equation}}
\newcommand{\ee}{\end{equation}}
\newcommand{\bea}{\begin{eqnarray}}
\newcommand{\eea}{\end{eqnarray}}

\title{Subleading properties of the QCD flux-tube in 3-d lattice gauge theory}

\ShortTitle{Subleading properties of the QCD flux-tube in 3-d lattice gauge theory}

\author{N.D. Hari Dass \thanks{Address after 26 Sep 2007: CHEP, Indian Institute of Science, Bangalore. 
Email:ndhari.dass@gmail.com}\\
        Hayama Center for Advanced Studies, Hayama, Japan\\
	E-mail: \email{hari@soken.ac.jp}}

\author{\speaker{Pushan Majumdar}\\
         Institut f\"ur Theoretische Physik, Westf\"alische-Wilhelms Universit\"at, M\"unster\\
         E-mail: \email{pushan@uni-muenster.de}}

\abstract{We
study the continuum limit of the string-like behaviour of flux tubes
formed between static quarks and anti-quarks in three dimensional $SU(2)$ lattice gauge theory.
We compare our simulation data with the predictions of both effective string models
as well as perturbation theory. On the string side we obtain clear evidence for
convergence of data to predictions of Nambu-Goto theory. We comment on the scales
at which the static potential starts
departing from one loop perturbation theory and then again being well
described by effective string theories. We
also estimate the leading corrections to the one-loop
perturbative potential as well as the  Nambu-Goto effective string.
In the intermediate regions we find an empirical formula which
gives surprisingly good fits.}

\FullConference{The XXV International Symposium on Lattice Field Theory\\
                   July 30 - August 4 2007\\
                     Regensburg, Germany}

\begin{document}

\section{Introduction}

Confinement of quarks, at least on the lattice seems to be due to the formation of a flux tube 
between a quark and an anti-quark in the QCD vacuum \cite{bali}. It has been conjectured that 
the properties of this flux tube can be described by an effective hadronic string \cite{effstring}.

On the lattice, this flux tube can be observed by measuring the potential between a static quark
 and an anti-quark. 
 One of the characteristics of the string like behaviour of the
 flux tube is the presence of a long distance $1/r$ term in the $q\bar q$ potential
 in all dimensions and with a universal coefficient. It is known as the L\"uscher term \cite{LSW,L}.

The L\"uscher term has been looked at in lattice simulations since the eighties \cite{ambjorn, deF}. 
However 
in recent times, increase in computing power and improvement in algorithms have allowed really 
precise measurements of that term and also the next subleading $r^{-3}$ term. See 
\cite{caselle, lw1, lw2,  4dim, pushspec1, HM5, kuticr} for example.

Another aspect of the string behaviour is the excitation spectrum of the flux tube. 
A recent review on this topic can be found in \cite{kutilat05}.

In this article we present results for the continuum limit of our simulations of the Polyakov loop
correlators for $d=3$ $SU(2)$ Yang-Mills theory and compare the resulting
static potential with both perturbation theory and string model
predictions. This allows us to narrow down bounds on 
the distance beyond which we can say the flux tube indeed shows a string like behaviour.

\section{Simulation parameters}
We carried out simulations of three dimensional $SU(2)$ lattice gauge theory on 
 lattices at four different lattice spacings.
On these lattices the scale was set by the Sommer parameter
$r_0=0.5~{\rm fm}$, implicitly defined
by $r_0^2 f(r_0)=1.65$, where $f(r)$ is the force between the static quark
and the anti-quark. Our measured values were $r_0/a=(3.9536(3), 6.2875(10), 
8.6602(8), 10.916(3))$ at $\beta =(5.0, 7.5, 10.0, 12.5)$ respectively. 
 We also observed that $\sigma r_0^2$ ($\sigma :$ string tension) is a constant 
$(\simeq 1.522)$ to a very good
 approximation as expected in the continuum limit.
 With this scale our coarsest lattice had a 
spacing of slightly below 0.13 fm and our finest lattice spacing is about 
0.045 fm. We used symmetric cubic lattices and the Wilson gauge action.
On all these
lattices, we computed Polyakov loop correlators
$\langle P^*(x)P(y)\rangle$ for various spatial separations $r=y-x$,
with separations being taken along the axes only.

To reliably extract signals of these observables which are exponentially
decreasing functions of $r$ and $T$ (the temporal extent of the lattice), 
we used the L\"uscher-Weisz exponential error reduction 
algorithm \cite{lw1}. For further details of our run parameters, see \cite{DM2}.

\section{Results}
From the $\langle P^*P\rangle$
correlator one can extract the static quark-antiquark potential $V(r)$ by
$
V(r)=-\frac{1}{T}\ln \langle P^*P(r)\rangle.
$ 
We prefer to look at the first and the second 
derivative of this potential for the force between the quark and the antiquark and information
 about sub-leading terms.
To facilitate our comparison with string models we
actually compute a scaled second derivative which we call $c(\tilde r)$. This quantity is expected to
become the L\"uscher term ($=-\frac{(d-2)\pi}{24}$) asymptotically.
\begin{figure}
\begin{center}
\includegraphics[width=0.8\textwidth,angle=0]{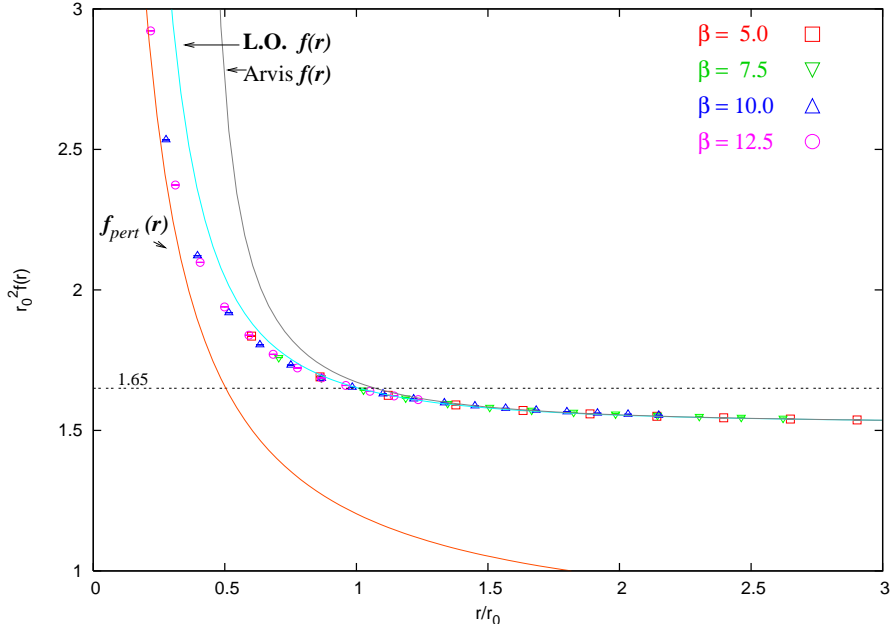}
\caption{
$r_0^2f(r)$ vs $r/r_0$. 
The horizontal line locates the Sommer scale. }
\label{force:fig}
\end{center}
\vspace*{-5mm}
\end{figure}

On the lattice these are given by
\be 
F({\bar r})=V(r)-V(r-1)\quad \& \quad
c({\tilde r})=\frac{{\tilde r}^3}{2}[V(r+1)+V(r-1)-2V(r)]
\ee
where ${\bar r}=r+\frac{a}{2}+{\cal O}(a^2)$ and ${\tilde r}=r+{\cal O}(a^2)$ are defined as 
in \cite{lw2} to reduce lattice artifacts.

For large distances, we are going to concentrate on the potential due 
to the NG string, the so called Arvis potential \cite{Ar} given by
$
V_{\rm Arvis}=\sigma r \left ( 1-\frac{(d-2)\pi}{12\sigma r^2}\right )^{1/2}.
$
We will define the {\bf L.O.} and {\bf N.L.O.} approximations by retaining $1/r$ and $1/r^3$ terms 
in the potential respectively.
 We will compare our lattice 
data on force and $c(r)$ with predictions from leading order, NLO and from the full Arvis potential. 

At short distances we compare with the perturbative potential obtained by Schr\"oder \cite{pert} as 
$
V_{\rm pert}(r)=s_{\rm pert} r + \frac{g^2C_F}{2\pi}\ln g^2r + \ldots 
$
with $s_{\rm pert}= \frac{7g^4C_FC_A}{64\pi}$. 
For $SU(2)$ $C_F=3/4$, $C_A=2$. 
The perturbative force and $c_{\rm pert}(r)$ can 
be computed by $f_{\rm pert}(r)=\frac{d V_{\rm pert}(r)}{d r}$ and $c_{\rm pert}(r)
=\frac{r^3}{2}\frac{d^2 V_{\rm pert}(r)}{d r^2}.$

In Fig. \ref{force:fig} we plot $r_0^2f(r)$ versus $r/r_0$. 
The horizontal line is $r^2f(r)=1.65$ and 
defines the Sommer scale $r_0$.
The {\bf N.L.O.} curve lies in between the {\bf L.O.} and Arvis curves and has 
been omitted for the sake of clarity. The data 
starts departing from the one-loop perturbative curve around 0.22 $r_0$ or 1.8 GeV and 
joins onto the string curves around 1.5 $r_0$ or 260 MeV. 
The scaling exhibited by the data is very good with all the four different beta values 
falling on the same curve. Beyond 1.5$r_0$ it is virtually 
impossible to distinguish the different theoretical curves as they are all dominated
 by the string tension. The force data in fact gives the impression of
the string description being good even at distances as small as $r_0$.

\begin{figure}
\begin{center}
\includegraphics[width=0.85\textwidth,angle=0]{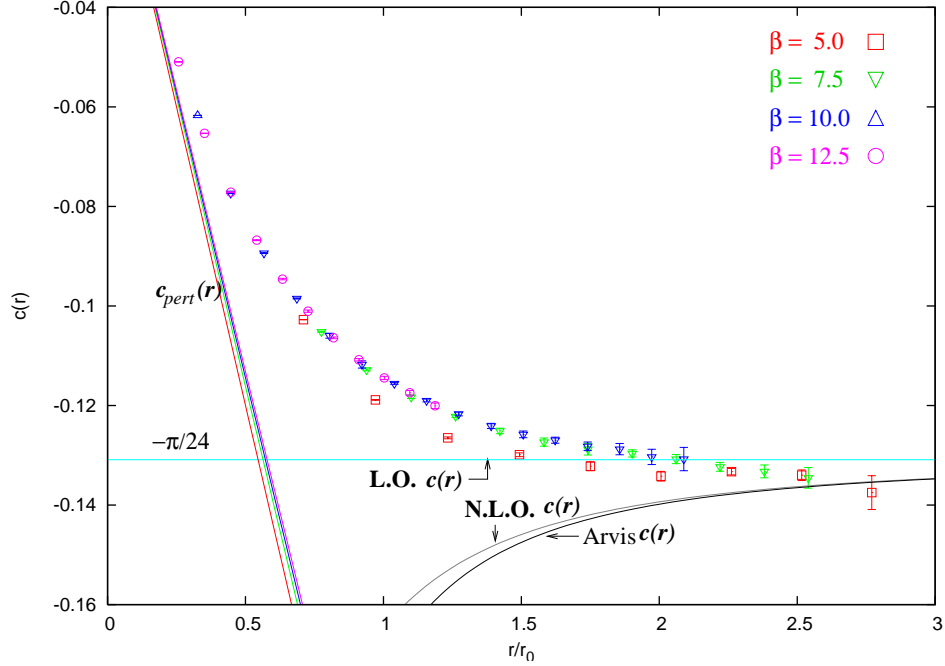}
\caption{$c(\tilde r)$. $c_{\rm pert}(r)$ : 1-loop perturbation theory. 
$\beta=12.5$ closest to data and $\beta=5$ farthest.}
\label{fig:cr}
\end{center}
\vspace*{-5mm}
\end{figure}

$c(\tilde r)$ does not contain the string tension, and has a universal
value in the L.O. It is therefore more sensitive to the 
sub-leading behaviour of the flux tube. In Fig. \ref{fig:cr} we plot 
$c(\tilde r)$ in units of $r/r_0$. 
We look at a wide range of $r$ starting from where the data almost touches the 
perturbative curves going all the way to the region where the string predictions hold.

\begin{table}[t]
\begin{center}
{\small
\begin{tabular}{r|r|l|l|l|l}
\hline
$r$ &  $\tilde r$~~ & ~~$\beta = 5 $ & ~~$\beta = 7.5 $ & ~~$\beta = 10 $ & ~~$\beta = 12.5 $ \\
\hline
 3&   2.808 & $-$0.10274 (3)&                &  $-$0.06172 (1)   &  $-$0.05095 (1) \\
 4&   3.838 & $-$0.11886 (8)&                &  $-$0.07737 (3)   &  $-$0.06530 (3) \\
 5&   4.876 & $-$0.1265 ~(2)&  $-$0.1052 (1) &  $-$0.08937 (6)   &  $-$0.07714 (5) \\
 6&   5.903 & $-$0.1299 ~(4)&  $-$0.1128 (2) &  $-$0.0985 ~(1)  &  $-$0.0868 ~(1)\\
 7&   6.920 & $-$0.1322 ~(9)&  $-$0.1183 (3) &  $-$0.1060 ~(5)  &  $-$0.0946 ~(2)\\
 8&   7.932 & $-$0.1342 (10)&  $-$0.1222 (3) &  $-$0.1118 ~(7)  &  $-$0.1011 ~(2)\\
 9&   8.941 & $-$0.1333 ~(6)&  $-$0.1251 (6) &  $-$0.1156 ~(2)  &  $-$0.1064 ~(2)\\
10&   9.948 & $-$0.1340 (10)&  $-$0.1273 (8) &  $-$0.1190 ~(3)  &  $-$0.1108 ~(3)\\
11&   10.95 & $-$0.1375 (34)&  $-$0.1288 (11)&  $-$0.1217 ~(4)  &  $-$0.1145 ~(4)\\
12&   11.96 &               &  $-$0.1296 (7) &  $-$0.1241 ~(5)  &  $-$0.1174 ~(5)\\
13&   12.96 &               &  $-$0.1308 (9) &  $-$0.1258 ~(7)  &  $-$0.1201 ~(6)\\
14&   13.96 &               &  $-$0.1323 (9) &  $-$0.1270 ~(6)  \\
15&   14.97 &               &  $-$0.1332 (13)&  $-$0.1282 ~(9)  \\
16&   15.97 &               &  $-$0.1345 (21)&  $-$0.1288 (11)  \\
17&   16.97 &               &                &  $-$0.1303 (15)  \\
18&   17.97 &               &                &  $-$0.1308 (23)  \\
\hline
\end{tabular}
}
\caption{$c(\tilde r)$}
\end{center}
\vspace*{-5mm}
\end{table}

The data almost lie on top of each other exhibiting nice scaling behaviour  
as one goes to larger values of $r$. The $\beta=12.5$ and $\beta=10$ data come together 
 already in the range 0.5 and 1.25 $r_0$. The $\beta=7.5$ set joins onto this 
at around 1.5$r_0$ and even the $\beta=5$ data joins up at around 2.25$r_0$. This points 
to the possibility that the continuum limit of the scale where the flux tube is well 
described by the Arvis curve can be obtained even on relatively coarse lattices. 

We estimate a temporal extent correction factor of about 0.1\% for $c(r)$ for our largest 
$r$ values. This is about an order of magnitude lower than our statistical errors at such $r$ values. 
Corrections due to finite spatial extents are of similar magnitude.

\begin{figure}[h]
\begin{center}
\includegraphics[width=0.85\textwidth,angle=0]{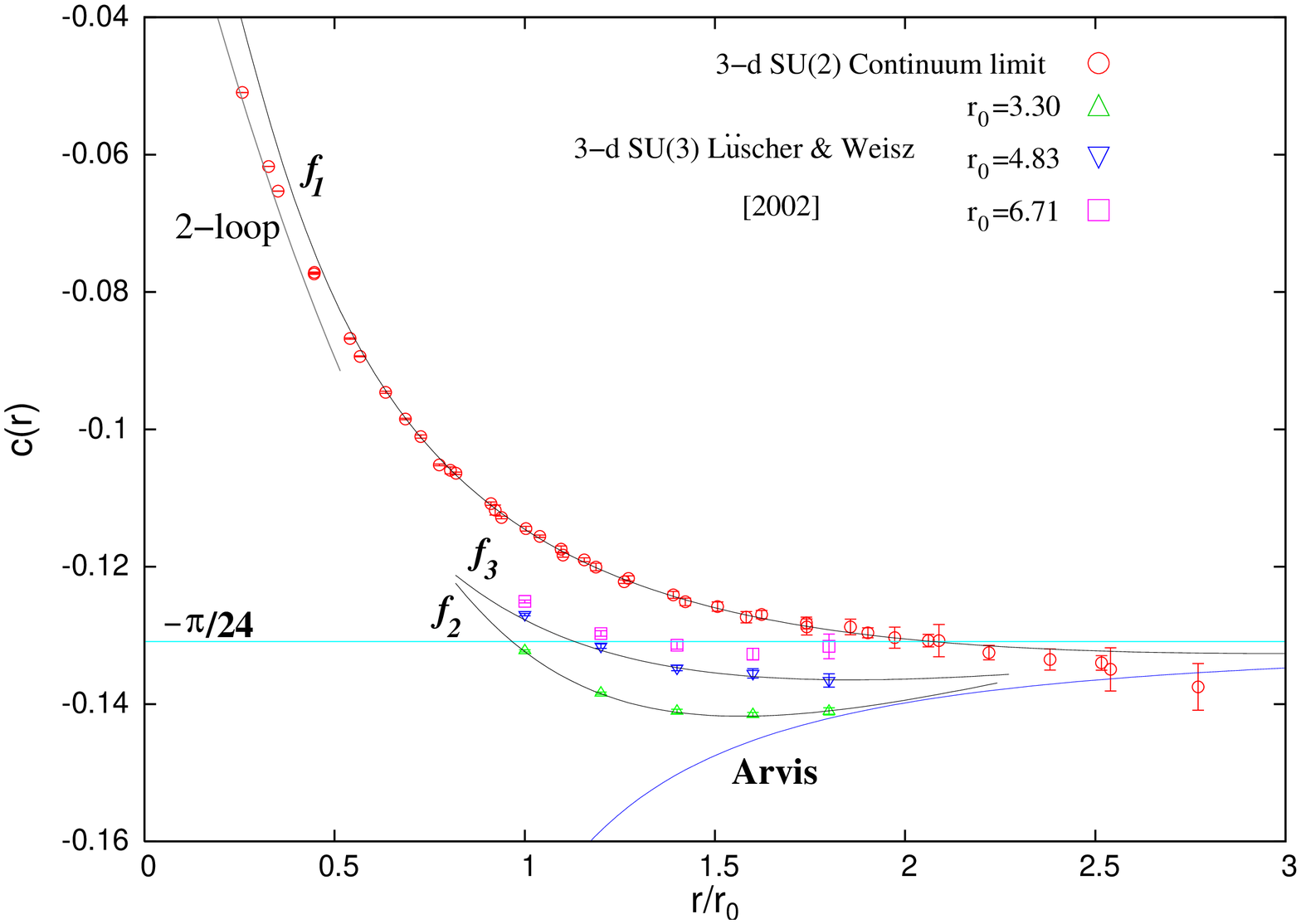}
\caption{ 
Curves of the type given in eqn. (\protect \ref{LJ}). 
$f_1$, $f_2$ and $f_3$ have $(a,b,n)$ of $(0.444, -0.258, 0.357)$ ,
$(0.458, -0.289, 0.691)$ and $(0.442, -0.287, 0.498)$. 
 2-loop :
$-\frac{g^2 r_0C_F}{4\pi} \frac{r}{r_0} +\frac{Ag^4r_0^2}{2}\left (\frac{r}{r_0}\right )^2. ~~~~g^2r_0=3.792$}
\label{LJ:fig}
\end{center}
\end{figure}

In $d=3$, IR divergences prevent computation of the perturbative
potential beyond one loop\footnote{NDH wishes to thank G. `t Hooft for an 
illuminating discussion on this.}. We tried to obtain these terms from our data. 
Assuming the perturbative potential to be of the form
\be
V_{\rm pert}(r) = \frac{g^2C_F}{2\pi} \ln g^2 r + \frac{7C_FC_Ag^4}{64\pi} r + A g^4r \ln g^2 r + Bg^6 r^2
+\ldots
\ee
we obtain $A=0.013162(3)$ and $B=0.001089(1)$.
Also we estimate the range of validity of first order perturbation theory to be about 
$ 0.1$ fermi (consistent with our estimate from the force data). 

 From our data we find that the string tension constitutes 95\% of the force at around 1.3$r_0$,
 98\% at around 2.1$r_0$ and 99\% at around 2.9$r_0$. The relative difference
 between the Arvis and the leading order force, which
 gives us an idea about the importance of the subleading behaviour, is about 2\%
 at 1.02$r_0$, 1\% at 1.2$r_0$ and goes down to 0.1\% at about 1.9$r_0$. 

The type of the string is determined by the sub-leading behaviour of the flux tube as 
at leading order, a variety of theories yield the universal L\"uscher term \cite{dietz,naik}.
It is clear from the data 
that the approach to the L\"uscher term is from below, consistent with effective bosonic string 
model predictions while at short distances the data matches perturbation theory.
Beyond a distance of about 2.75$r_0$, the data seems to be well described by the Arvis curve. 

An interesting question is what happens to this distance for $SU(N)$ as $N$ increases.
For $SU(3)$ on coarsest lattices \cite{lw2} it is about 1.8$r_0$. Similar indications 
are also there for $SU(5)$ \cite{HM5}.
However this scale shifts towards larger $r$ as one approaches the continuum limit. 

Beyond {\bf N.L.O.}, effective string theories parametrise $c(r)$ to be of the type
$\alpha(r_0/r)^4 + \beta(r_0/r)^6$. Our data gives $\alpha=0.209(9)$ and $\beta=-0.235(24)$, which is 
quite different from the Arvis values. It is therefore interesting to know the effective 
string predictions for these coefficients.

At intermediate distances over a wide region of $r$ varying from $0.5r_0$ to $2.8r_0$, the data 
is very well described by a formula of the type 
\be\label{LJ}
c(\tilde r)= a\left (\frac{1}{x^{2n}}-\frac{1}{x^n}+\frac{b}{x^{3n}}\right ).
\ee  
Existing $SU(3)$ data in 3-d \cite{lw2} also admit a similar description.
The curves are shown in Fig. \ref{LJ:fig}. It is not clear if there is any theoretical 
basis for such a description, but at least they provide accurate interpolation formulae. 

\section{Conclusions}

In this article we have looked at the continuum limit behaviour of the $SU(2)$ flux 
tube at intermediate distances by measuring the static $q\bar q$ potential.
 Starting from a distance of about 0.1 fermi where the potential starts 
breaking away from 1-loop perturbation theory, we go to distances of about 1.4 fermi 
where the data begins to be well described by the Arvis potential. 

Our data on $c(\tilde r)$ seems to approach the L\"uscher term from below as 
expected in bosonic string models. At distances below 0.15 fermi the data joins onto 
the perturbative values. An empirical formula describes the 
data well in the intermediate region. 

\acknowledgments
One of the authors, PM, gratefully acknowledges the numerous discussions 
 with Peter Weisz. NDH thanks the Hayama Centre for Advanced Studies for
hospitality. The simulations
were carried out on the teraflop Linux cluster KABRU at IMSc as part of the Xth plan project ILGTI.

\end{document}